\documentclass[aps, prd, floatfix, nofootinbib, superscriptaddress, twocolumn]{revtex4-1}

\usepackage{latexsym}
\usepackage{amsmath}
\usepackage{amssymb}
\usepackage{amsfonts}
\usepackage{textcomp}
\usepackage{color}

\usepackage[mathscr,scaled=1.15]{urwchancal}
\DeclareFontFamily{OT1}{pzc}{}
\DeclareFontShape{OT1}{pzc}{m}{it}%
{<-> s * [1.15] pzcmi7t}{}
\DeclareMathAlphabet{\mathpzc}{OT1}{pzc}{m}{it}

\usepackage{color}

\usepackage{supertabular}
\usepackage{placeins}
\usepackage{epsfig}
\usepackage{graphicx}

\definecolor{purple}{rgb}{0.5,0,0.5}
\definecolor{blue}{rgb}{0.0,0,0.9}
\definecolor{prdblue}{rgb}{0.133,0.118,0.498}
\usepackage[colorlinks=true, pdfstartview=FitV, linkcolor=prdblue, citecolor= prdblue, urlcolor=prdblue]{hyperref}

\def\s{\scriptscriptstyle}

\makeatletter

\setbox0\hbox{$\xdef\scriptratio{\strip@pt\dimexpr
    \numexpr(\sf@size*65536)/\f@size sp}$}

\newcommand{\scriptveryshortarrow}[1][3pt]{{%
    \hbox{\rule[\scriptratio\dimexpr\fontdimen22\textfont2-.2pt\relax]
               {\scriptratio\dimexpr#1\relax}{\scriptratio\dimexpr.4pt\relax}}%
   \mkern-4mu\hbox{\let\f@size\sf@size\usefont{U}{lasy}{m}{n}\symbol{41}}}}

\makeatother

\begin{document}


\title{$\,$\\[-6ex]\hspace*{\fill}{\normalsize{\sf\emph{Preprint no}.\ NJU-INP 033/21}}\\[1ex]
Fresh extraction of the proton charge radius from electron scattering}

\date{2021 July 18}

\author{Zhu-Fang Cui}
\affiliation{School of Physics, Nanjing University, Nanjing, Jiangsu 210093, China}
\affiliation{Institute for Nonperturbative Physics, Nanjing University, Nanjing, Jiangsu 210093, China}
\author{Daniele Binosi}
\email{binosi@ectstar.eu}
\affiliation{European Centre for Theoretical Studies in Nuclear Physics
and Related Areas; Villa Tambosi, Strada delle Tabarelle 286, I-38123 Villazzano (TN), Italy}
\author{Craig D.~Roberts}
\email[]{cdroberts@nju.edu.cn}
\affiliation{School of Physics, Nanjing University, Nanjing, Jiangsu 210093, China}
\affiliation{Institute for Nonperturbative Physics, Nanjing University, Nanjing, Jiangsu 210093, China}
\author{Sebastian M.~Schmidt}
\affiliation{Helmholtz-Zentrum Dresden-Rossendorf, Dresden D-01314, Germany}
\affiliation{RWTH Aachen University, III. Physikalisches Institut B, Aachen D-52074, Germany}

\begin{abstract}
We present a novel method for extracting the proton radius from elastic electron-proton ($ep$) scattering data. The approach is based on interpolation via continued fractions augmented by statistical sampling and avoids any assumptions on the form of function used for the representation of data and subsequent extrapolation onto $Q^2\simeq 0$.  Applying the method to extant modern $e p$ data sets, we find that all results are mutually consistent and, combining them, arrive at $r_p=0.847(8)\,$fm.
This result compares favourably with values obtained from contemporary measurements of the Lamb shift in muonic hydrogen, transitions in electronic hydrogen, and muonic deuterium spectroscopy.
\end{abstract}

\maketitle


\noindent\emph{1.$\;$Introduction} ---
The proton is Nature's most fundamental bound state. Composed of three valence constituents, two $u$-quarks and one $d$-quark, it seems to be absolutely stable: in the $\sim$ 14-billion years since the Big Bang, proton decay has not been observed. The proton’s extraordinarily long lifetime is basic to the existence of all known matter. Yet, the forces responsible for this remarkable feature are not understood.

Proton structure is supposed to be described by quantum chromodynamics (QCD), the Standard Model quantum field theory intended to explain the character and interactions of the proton (and all related objects) in terms of gluons (gauge fields) and quarks (matter fields) \cite{Marciano:1979wa}. Today, the proton's mass, $m_p$, can be calculated with good accuracy using modern theoretical tools \cite{Durr:2008zz, Eichmann:2016yit, Qin:2019hgk}; but that is not the case for its radius, $r_p$.

The proton's radius is of particular importance because it relates to the question of confinement, \emph{viz}.\ the empirical fact that no isolated gluon or quark has ever been detected.  The value of $r_p$ characterises the size of the domain within which the current-quarks in QCD's Lagrangian may rigorously be considered to represent the relevant degrees of freedom.  (A clearer notion of confinement may appear in a proof that quantum SU$_c(3)$ gauge field theory is mathematically well-defined, \emph{i.e}.\ a solution to the Yang-Mills ``Millennium Problem'' \cite{millennium:2006}.)  Moreover, it is not just strong interactions which feel the size of $r_p$.  An accurate value of the proton's charge radius is also crucial to a precise determination of quantities in atomic physics, such as the Rydberg constant and Lamb shift.

\begin{figure}[!b]
	\includegraphics[width=0.99\linewidth]{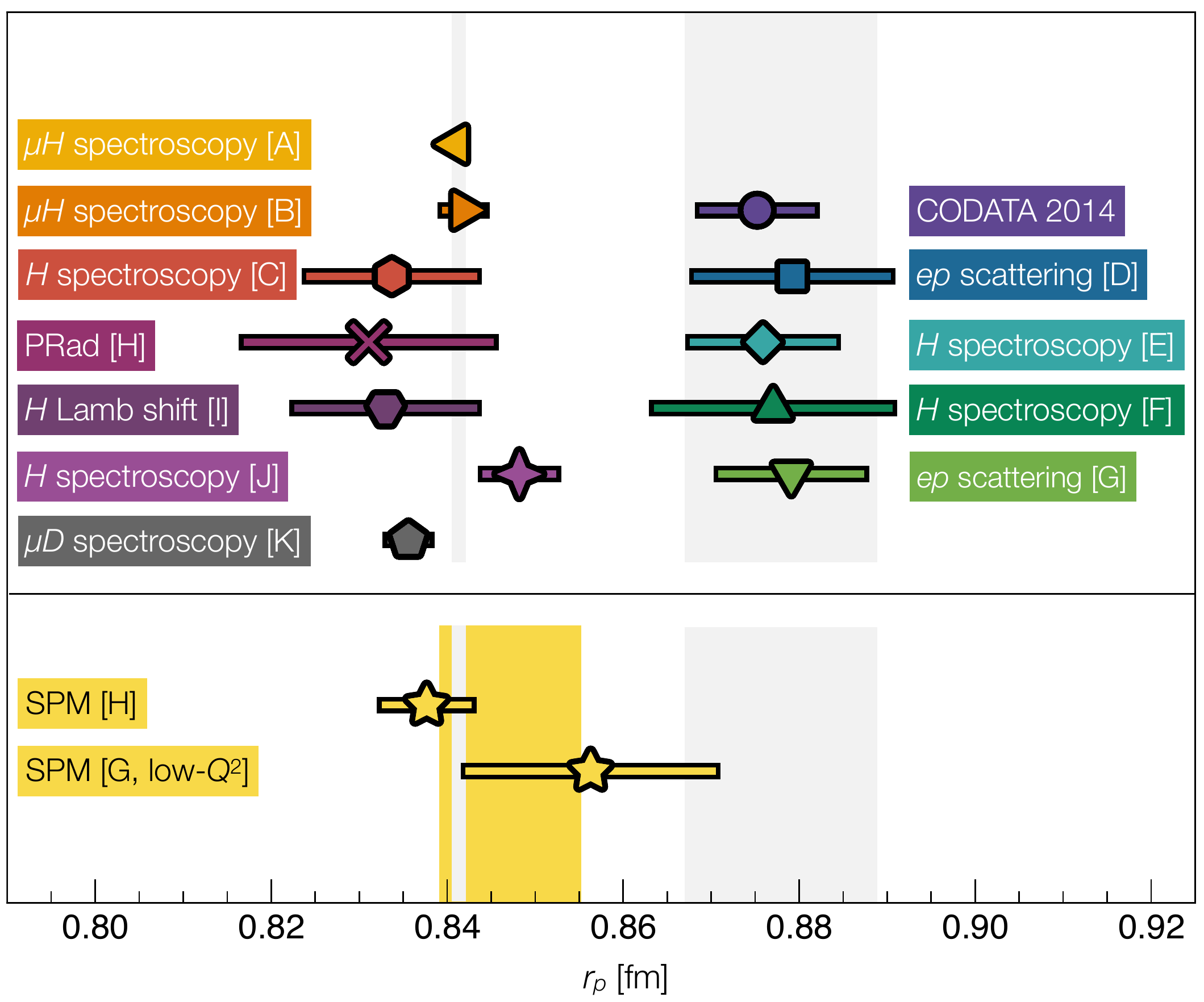}
	\caption{\label{fig:worldav}
{\it Upper panel}.  $r_p$ measurements, various techniques:
CODATA = Ref.\,\cite{Mohr:2015ccw};
[A] = Ref.\,\cite{Antognini:1900ns};
[B] = Ref.\,\cite{Pohl:2010zza};
[C] = Ref.\,\cite{Beyer:2017gug};
[D] = $ep$ scattering average from Ref.\,\cite{Mohr:2015ccw};
[E] = $H$ spectroscopy average from Ref.\,\cite{Mohr:2015ccw};
[F] = Ref.\,\cite{Fleurbaey:2018fih};
[G] = Ref.\,\cite{Bernauer:2010wm};
[H] = Ref.\,\cite{Xiong:2019umf};
[I] = Ref.\,\cite{Bezginov1007};
[J] = Ref.\,\cite{Grinin1061};
and [K] muonic deuterium spectroscopy from Ref.\,\cite{Pohl1:2016xoo}.
{\it Lower panel}. Results obtained from the data in Refs.\cite{Bernauer:2010wm, Xiong:2019umf} using the Schlessinger Point Method (SPM) \cite{PhysRev.167.1411, Schlessinger:1966zz, Tripolt:2016cya} as described herein.}
\end{figure}

Naturally, $m_p$ and $r_p$ are correlated.  A solution to the Standard Model will deliver values for both.  Hence, precise measurements are necessary to set rigorous benchmarks for theory.  The problem is that whilst the relative error on $m_p$ is $\sim10^{-10}$, measurements of $r_p$ now disagree amongst themselves by as much as eight standard deviations, $8\sigma$, as illustrated in Fig.\,\ref{fig:worldav}\,--\,upper panel.  This conflict, which emerged following extraction of the proton radius from measurements of the Lamb shift in muonic hydrogen ($\mu H$) \cite{Pohl:2010zza}, has come to be known as the ``proton radius puzzle'' \cite{Pohl:2013yb,Carlson:2015jba}.

Many solutions of this puzzle have been offered, \emph{e.g}.\
some unknown QCD-related corrections may have been omitted in the muonic hydrogen analysis and their inclusion might restore agreement with the electron-based experiments that give a larger value;
the discrepancy could signal some new interaction(s) or particle(s) outside the Standard Model, which lead to a violation of universality between electron ($e$) and muon ($\mu$) electromagnetic interactions;
or some systematic error(s) has (have) hitherto been neglected in the analysis of electron scattering.

Empirically, novel experiments have been proposed in order to test various possibilities, including $\mu p$ elastic scattering (MUSE) \cite{Gilman:2013vma} and $e p$ scattering at very low momentum transfer (PRad) \cite{Gasparian:2014rna}.  PRad recently released its result~\cite{Xiong:2019umf}:
\begin{align}
	r^\mathrm{PRad}_p=0.831\pm0.007_\mathrm{stat}\pm0.012_\mathrm{syst}\ [\mathrm{fm}].
	\label{PRad-res}
\end{align}
Significantly, this is the first published analysis of an $ep$ scattering experiment to obtain a result in agreement with the radius extracted from $\mu H$ measurements.

In performing and analysing the $e p$ scattering experiment, the PRad collaboration implemented a number of improvements over previous efforts, which included:
reaching the lowest yet achieved momentum-transfer-squared, $Q^2 = 2.1\times 10^{-4}\,$GeV$^2$;
and covering an extensive domain of low $Q^2$: $2.1\times 10^{-4} \leq Q^2/{\rm GeV}^2\leq 6 \times 10^{-2}$.
Moreover, since the charge radius is obtained as
\begin{align}
r_p^2 = \left. - \frac{6}{G_E^p(0)} \frac{d}{dQ^2} G_E^p(Q^2)\right|_{Q^2=0},
\label{rp_def}	
\end{align}
where $G^p_E(Q^2)$ is the proton's elastic electromagnetic form factor, PRad paid careful attention to the impact of the choice of fitting form on the extracted charge radius, an issue highlighted previously \cite{Kraus:2014qua, Lorenz:2014vha, Griffioen:2015hta, Higinbotham:2015rja, Hayward:2018qij, Zhou:2018bon, Alarcon:2018zbz, Higinbotham:2019jzd, Hammer:2019uab}.
Notably, their functional form was predetermined through a bootstrap procedure applied to pseudodata generated with fluctuations mimicking the $Q^2$-binning and statistical uncertainty of the experimental setup, \emph{i.e}.\ without knowledge of the actual PRad data \cite{Yan:2018bez}.  While this procedure renders the PRad extraction robust, it also means that, ultimately, a specific functional form was chosen \cite{Yan:2018bez}.

We reanalyse the PRad data \cite{Xiong:2019umf} and also data from the A1 Collaboration \cite{Bernauer:2010wm} using a statistical Schlessinger Point Method (SPM) \cite{PhysRev.167.1411, Schlessinger:1966zz}.
Following Ref.\,\cite{, Tripolt:2016cya}, the SPM has been used widely and effectively to solve numerous problems in hadron physics, especially those which demand model-independent interpolation and extrapolation, \emph{e.g}.\ Refs.\,\cite{Chen:2018nsg, Binosi:2018rht, Binosi:2019ecz, Eichmann:2019dts, Yao:2020vef}.
In this approach, no functional form is assumed.
Instead, one arrives at a set of continued fraction interpolations capable of capturing both local and global features of the curve that the data are supposed to be measuring.  This latter aspect is crucial because it ensures that the validity of the constructed curves extends outside the data range limits, ultimately allowing for the evaluation of the curves' first derivative at the origin.  A robust estimation of the error is also obtained by means of a statistical bootstrap procedure \cite{10.5555/1403886}.

\smallskip

\noindent\emph{2.$\;$Theory for interpolation and extrapolation of smooth functions} ---
The foundation for our fresh analysis of $G_E^p(Q^2)$ data, obtained from $ep$ scattering and available on $Q_{\rm min}^2 \leq Q^2 \leq Q_{\rm max}^2$, is the SPM.
In general, given $N$ pairs, ${\mathsf D} = \{(x_i,y_i=f(x_i))$\}, being the values of some smooth function, $f(x)$, at a given set of discrete points, a basic SPM application constructs a continued-fraction interpolation:
\begin{equation}
{\mathpzc C}_N(x) = \frac{y_1}{1+\frac{a_1(x-x_1)}{{1+\frac{a_2(x-x_2)}{\vdots a_{N-1}(x-x_{N-1})}}}}\,,
\end{equation}
in which the coefficients $\{a_i|i=1,\ldots, N-1\}$ are constructed recursively and ensure ${\mathpzc C}_N(x_i) = f(x_i)$, $i=1\,\ldots,N$.  The SPM is related to the Pad\'e approximant; and the procedure accurately reconstructs any analytic function within a radius of convergence fixed by that one of the function's branch points which lies closest to the domain of real-axis points containing the data sample.  For example, suppose one considers a monopole form factor represented by $N>0$ points, each one lying on the curve; then using any one of those points, the SPM will exactly reproduce the function.

In the physical cases of interest herein, one deals with data that are distributed statistically around a curve for which the SPM must deliver an accurate reconstruction.  Given that all sets considered are large, $N$ is big enough to enable the introduction of a powerful statistical aspect to the SPM.
Namely, one randomly selects $M < N$ points from the set $\mathsf{D}$, typically with $4 < M \lesssim N/2$ \cite{Chen:2018nsg, Binosi:2019ecz}.  In theory, one can then obtain $C(N,M)$ different interpolating functions; in practice, this number is reduced by introducing physical constraints on their behavior. The minimal $N$ we consider is $N=33$, \emph{i.e}.\ the PRad data set at a beam energy of $1.1\,$GeV; thus, choosing $M \in [6,17]$ gives ${\cal O}(10^6 - 10^9)$ possible interpolators, out of which we select the first $5\times 10^3$ corresponding to smooth monotonic functions on the entire $Q^2$ domain.  No further restriction is imposed; specifically, no unity constraint on $G_E^p(Q^2=0)$ is required.

Each interpolating function defines an extrapolation to $Q^2=0$, from which $r_p$ can be extracted using Eq.\,\eqref{rp_def}.  For a given value of $M$, the value of the radius is then obtained as the average of all results obtained from the $5\,000$ curves.

To estimate the error associated with the SPM determined proton radius, one needs to account for the experimental errors in each of the data sets.  This can be achieved by using a statistical bootstrap procedure.  To wit, we generate $1\,000$ replicas for each set by replacing each datum by a new point, randomly distributed around a mean defined by the datum itself with variance equal to its associated error.
The probability distribution function, ${\mathpzc N}(\mu,\sigma)$, characterising the $r_p$ values extracted via the procedure described above on each replica is, to a very good approximation, normal, with average $r^{\s M}_p$, standard deviation $\sigma=\sigma_{r}^{\s M}$, skewness $\beta^{\s M}_3\approx0$ and kurtosis $\beta^{\s M}_4\approx3$ (see Supplemental Material, Fig.\,S1).
In addition, the fact that $M$ is not fixed leads to a second error source $\sigma_{\s{\delta\!M}}$, which can be estimated by changing $M\to M'$, repeating the aforementioned procedure for this new $M'$-value, and evaluating the standard deviation of the distribution of $r^{\s M}_p$ for different $M$ values.

Consequently, for each kinematics, the SPM result is
\begin{align}
	&r_p\pm\sigma_r;
	&r_p=\sum_{j=1}^{n_{\s M}}\frac{r^{\s{M_j}}_p}{n_{\s{M}}};&
&\sigma_r=\Bigg[\sum_{j=1}^{n_M}\frac{(\sigma^{\s{M_j}}_r)^2}{n^2_{\s{M}}}
+\sigma_{\s{\delta\!M}}^2\Bigg]^{\frac12}.
	\label{SPMrp}
\end{align}
Herein, we compute results for each one of the values $\{M_j=5+j\,\vert\ j = 1,\dots,n_{\s{M}};\,n_{\s{M}} = 12\}$, so that for any given data set we have 60-million 
values of $r_p$, each calculated from an independent interpolation; and, typically, we find $\sigma_{\s{\delta\!M}}\ll \sigma_{r}^{\s M_j}$ for all $j$s in the range specified above.  (See Supplemental Material, Eq.\,II.1).

\smallskip

\noindent\emph{3.$\;$Smoothing with roughness penalty} --- Before implementing the statistical SPM, however, one issue must be addressed.  Namely, as highlighted above, sound experimental data are statistically scattered around that curve which truly represents the observable.  They do not lie on the curve; hence, empirical data should not be directly interpolated.

A solution to this problem is \emph{smoothing} with a \emph{roughness penalty}, an approach we have implemented following the ALGOL procedure detailed in \cite{Reinsch:1967aa} and which we now sketch.
One begins by assuming the data are good, \emph{viz}.\ they are a true measurement of an underlying smooth function.
The next step is to identify the correct basis functions for the smoothing operation.  These are provided by cubic splines, defined as follows.
Consider a sequence of increasing numbers $x_1 <x_2 <\dots<x_\ell$ in some interval ${\mathpzc I}=[a,b]$, $a < b$; and call these numbers knots.
A function $g$ defined on ${\mathpzc I}$ is a cubic spline with-respect-to (wrt) the knots $\{x_i\}$ if the following two conditions are satisfied.  (\emph{i}) $g$ is a cubic polynomial on each of the $m +1$ subintervals: $g(x) = a_i + b_i x + c_i x^2 + d_i x^3$, $x\in[x_i,x_{i+1}]$, $x_0=a$, $x_{\ell+1}=b$.
(\emph{ii}) $g$ is a ${\mathbb C}^2$ function, \emph{viz}.\ continuous with two continuous derivatives.
All cubic splines (wrt knots $\{x_i\}$) form a vector space of functions with $\ell + 4$ degrees-of-freedom.  A set of basis functions for this space is: $h_1(x)=1;\ h_2(x)=x;\ h_3(x)=x^2;\dots;\ h_{i+4}(x) = (x - x_i)^3_+$, where ``+'' means $x_i < x < x_{i+1}$.
A cubic spline on $[a, b]$ is called ``natural'' if its second and third derivatives vanish at the interval’s endpoints.

Now consider the Sobolev space ${\mathbb S}[{\mathpzc I}]$ of ${\mathbb C}^2$ functions on ${\mathpzc I}$.  In the roughness-penalty approach to smoothing, one seeks that function $g\in {\mathbb S}[{\mathpzc I}]$ which minimises
 \begin{align}
 	\mathsf{P}(g,\lambda)
 =\lambda\sum_{i=1}^\ell[y_i-g(x_i)]^2+(1-\lambda)\int^b_a\!\mathrm{d}x\,[g^{\prime\prime}(x)]^2.
 	\label{swrp}
 \end{align}
The first term in Eq.~\eqref{swrp} quantifies the data-fidelity of $g$.  The second term introduces the roughness penalty via the smoothing parameter $\lambda\in [0, 1]$: the original data are recovered for $\lambda\to  1$ (no penalty) and a linear least-squares fit is obtained as $\lambda\to 0$ (maximum penalty).

$\mathbb{S}[I]$ is an infinite-dimensional space; but it can be shown that when $\ell > 1$, the minimiser lies in the finite dimensional space of natural cubic splines with knots located at the $\ell$ data points, $\{x_i\}$. In fact, the following theorem holds.  Let $g$ be any smooth function on ${\mathpzc I}$ for which $g(z_i) = y_i,\ i = 1,\dots,\ell$; and suppose that $s$ is the natural cubic spline interpolant for the values $\{y_i\}$ at $\{x_i\}$; then
\begin{align}
	\int^b_a\!\mathrm{d}z\,[g^{\prime\prime}(z)]^2\geq \int^b_a\!\mathrm{d}z\,[s^{\prime\prime}(z)]^2,
\end{align}
with equality if and only if $g \equiv s$.


At this point the smoothing parameter $\lambda$ is somewhat arbitrary, with a typical value near $1/(1+h^3/6)$ where $h$ is the average spacing of the data sites:
\mbox{$h\sim{5\times 10^{-4},1.5\cdot10^{-3}},$} and $1\times 10^{-3}$ for the PRad data at beam energy $1.1\,$GeV, $2.2\,$GeV and their combination \cite{Xiong:2019umf};
and $h\sim1.7\times 10^{-3}$ for the Mainz data \cite{Bernauer:2010wm}.
On the other hand, an estimate of the \emph{optimal} value for $\lambda$ can be determined by means of a (generalised) cross-validation procedure \cite{Craven:1978aa}, which we now explain.  Pretend that observation ``$k$'' is lost, so that only the remaining $\ell-1$ points are available for constructing a smoothing spline with respect to $\lambda$.  Denote the solution of this reduced problem by $\check{s}_k$; by definition, $\check{s}_k$ minimises
\begin{align}
\lambda\sum_{i\neq k}^\ell[y_i-g(x_i)]^2+(1-\lambda)\int^b_a\!\mathrm{d}x\,[g^{\prime\prime}(x)]^2.	
\end{align}
The quality of $\check{s}_k$ as a predictor for a new observation can be judged by the difference $y_k - \check{s}_k(x_k)$; and this leads to the cross-validation procedure, \emph{i.e}.\ $\lambda_{\mathrm{opt}}$ is the value of $\lambda$ for which the following function is minimised:
\begin{equation}
	\mathsf{S}(\lambda)=\frac1\ell\sum_{i=1}^\ell[y_i-\check{s}_i(x_i)]^2.
\end{equation}
On average, we find $\lambda_\mathrm{opt}=1-\epsilon$, with $\epsilon\sim 2 \times 10^{-6}$, $6 \times 10^{-6}$, and $4 \times 10^{-6}$ for PRad $1.1\,$GeV beam energy, $2.2\,$GeV, and combined values, respectively, and $\epsilon\sim 1.5 \times 10^{-6}$ for the Mainz data.

\smallskip

\noindent\emph{4.$\;$Final procedure, validation and results} ---
As explained above, the SPM extraction of the proton radius from a set of $ep$ scattering data requires the following steps:
(\emph{i}) generate $1\,000$ replicas for the given experimental central values and uncertainties;
(\emph{ii}) smooth each replica with the associated optimal parameter $\lambda_{\mathrm{opt}}$;
(\emph{iii}) for each number of input points $M_j \in [6,17]$, determine the distribution of proton radii $r^{\s{M_j}}_p$, its associated $\sigma^{\s{M_j}}_r$, and the overall $\sigma_{\s{\delta\!M}}$;
and
(\emph{iv}) combine this information to obtain the final result for the proton radius and (statistical) error through Eq.\,\eqref{SPMrp}.

One might wonder if the proposed SPM extraction method is robust, \emph{i.e}.\ whether or not it can reliably extract the proton radius in a diverse array of cases.
We checked this by using a wide variety of models that have been employed to fit the world's $ep$ scattering data \cite{Borkowski:1975ume, Kelly:2004hm, Arrington:2003qk, Arrington:2006hm, Bernauer:2013tpr, Ye:2017gyb, Alarcon:2017ivh} to generate a proton electromagnetic form factor $G^E_p$ with a known value for the radius.
From these, we generated replicas with the $Q^2$-binning and uncertainties of the PRad \cite{Xiong:2019umf} and A1 \cite{Bernauer:2010wm} data sets.
In all cases, regardless of the generator employed, we found that the SPM returns the radius value used to generate the pseudodata; and, furthermore, the result is practically independent of the number of initial input points $M_j$. (See Supplemental Material, Sec.\,II.2).

The first $G^E_p$ data from which we extracted the proton radius are those from the PRad experiment \cite{Xiong:2019umf}, which reported data using $1.1\,$GeV ($N=33$) and $2.2\,$GeV $(N=38)$ electron beams.  Analysed separately, the SPM gives: $r^{1.1}_p/{\rm fm}=0.842\pm0.008_{\mathrm{stat}}$ for the $1.1\,$GeV data; and $r^{2.2}_p/{\rm fm}=0.824\pm0.003_{\mathrm{stat}}$ for the $2.2$\,GeV data.
Treated alone, the PRad data at $2.2\,$GeV leads to a lower value of the proton radius and a smaller error (one-third the size) than are obtained from the $1.1\,$GeV data; moreover, it drives the error in the PRad combined binning, reducing it to roughly one-half the value obtained using the $1.1\,$GeV data alone.  These observations accord with those made by the PRad Collaboration \cite{Xiong:2019umf}: see, in particular, Fig.\,S16 in the associated Supplemental Material.

Our final result, obtained from a combined analysis of the PRad data, is
\begin{equation}
	r^{{\mathrm{PRad}}}_p=0.838\pm0.005_{\mathrm{stat}}\ [\mathrm{fm}],
	\label{SPM-Prad}
\end{equation}
which is displayed in Fig.\,\ref{fig:worldav}\,--\,lower panel and reproduces, within errors, the published PRad result.

Data obtained in experiments performed by the A1 collaboration at Mainz \cite{Bernauer:2010wm} comprise $1\,400$ cross-sections measured at beam energies of $0.18$, $0.315$, $0.45$, $0.585$, $0.72$, and $0.855\,$GeV.  This collection of data stretches toward low $Q^2$, albeit not reaching the PRad values: $3.8 \times 10^{-3}\,$GeV$^2$ \emph{cf}.\ $2.1 \times 10^{-4}\,$GeV$^2$.  Therefore, we also applied our method to the A1 data, first restricting the analysis to the low-$Q^2$ region, consisting of $N=40$ data in the interval $3.8 \times 10^{-3} \leq Q^2/{\rm GeV}^2 \leq 1.4\times10^{-2}$.
In this case we obtained (Fig.\,\ref{fig:worldav}\,--\,lower panel):
\begin{align}
	r^{{\mathrm{A1}-\mathrm{low}Q^2}}_p=0.856\pm0.014_{\mathrm{stat}}\ [\mathrm{fm}].
	\label{SPM-A1}
\end{align}
Eliminating the restriction to the low-$Q^2$ region yields the same central value but a larger error:
$r^{{\mathrm{A1}}}_p/{\rm fm}=0.857\pm0.021_{\mathrm{stat}}$.
In this case, $\sigma_{\s{\delta\!M}}\sim \sigma_{r}^{\s M_j}$; so, extending the range of squared momentum transfer up to $Q^2\sim 1\,$GeV$^2$ limits the ability of the SPM to provide an $M$-independent result.
		
The original A1 Collaboration estimate is \cite{Bernauer:2010wm}: $r^{\mathrm{A1}-\mathrm{coll.}}_p/{\rm fm} =0.879\pm0.005_\mathrm{stat}\pm0.006_\mathrm{syst}$.  Thus, whilst the SPM reanalysis of the A1 data, Eq.\,\eqref{SPM-A1}, has a larger statistical uncertainty, it yields a value that agrees with both the PRad estimate and the $\mu H$ experiments.

\smallskip

\noindent\emph{5.$\;$Conclusions} ---
We calculated the proton charge radius, $r_p$, by analysing high-precision $ep$ scattering data obtained in modern experiments \cite{Xiong:2019umf, Bernauer:2010wm} using a statistical sampling approach based on the Schlessinger Point Method for the interpolation and extrapolation of smooth functions.  An important feature of this scheme is that no specific functional form is assumed for the interpolator, \emph{i.e}.\ it produces a form-unbiased interpolation as the basis for a well-constrained extrapolation.  All considered $ep$ scattering data sets yielded consistent results [Eqs.\,\eqref{SPM-Prad} and \eqref{SPM-A1}]; and combining them we find:
\begin{align}
	r^\mathrm{SPM}_p=0.847\pm 0.008_\mathrm{stat}\ [\mathrm{fm}],
\end{align}
which is indicated by the gold band in the lower panel of Fig.\,\ref{fig:worldav}.

Consequently, according to this analysis, there is no discrepancy between the proton radius obtained from $ep$ scattering and that determined from the Lamb shift in muonic hydrogen -- $r_p = 0.84136(39)\,$fm \cite{Pohl:2010zza, Antognini:1900ns}, the modern measurement of the $2S$\textrightarrow$4P$ transition-frequency in regular hydrogen -- $r_p = 0.8335(95)\,$fm \cite{Beyer:2017gug}, the Lamb shift in atomic hydrogen -- $r_p = 0.833(10)\,$fm \cite{Bezginov1007}, the combination of the latest measurements of the $1S$\textrightarrow$3S$ and $1S$\textrightarrow$2S$ transition frequencies in atomic hydrogen -- $r_p=0.8482(38)\,$fm \cite{Grinin1061}, or even the muonic deuterium determination $r_p=0.8356(20)\,$fm \cite{Pohl1:2016xoo}.
Furthermore, our analysis suggests that the explanation for the mismatch which spawned the ``proton radius puzzle'' lies in an underestimation of the systematic error introduced by the use of specific, limiting choices for the functions employed to interpolate and extrapolate $ep$ scattering data.

\smallskip

\noindent\emph{$\;$Acknowledgments} --- We are grateful for constructive comments from H.~Gao, D.\,W.~Higinbotham, R.\,J.~Holt, P.\,E.~Reimer and W.~Xiong.  Use of the computer clusters at the Institute for Nonperturbative Physics at Nanjing University is gratefully acknowledged.
Work supported by:
National Natural Science Foundation of China (under Grant No.\,11805097);
Jiangsu Provincial Natural Science Foundation of China (under Grant No.\,BK20180323);
and
Jiangsu Province \emph{Hundred Talents Plan for Professionals}.

\smallskip

\hspace*{-\parindent}\dotfill\ \emph{Supplemental Material} \dotfill

\medskip

\noindent\emph{SM\,I.$\;$Validation models and procedure.} ---
The validity of the SPM procedure for extracting the proton radius can be checked against replica data sets built from values of the proton elastic electromagnetic form factor, $G^p_E$, evaluated at the experimentally available $Q^2$ and generated using specific models in which the proton radius $r_p$ is known \emph{a priori}. To mimic the variability of real data, these values are then redistributed by introducing fluctuations drawn according to a normal distribution.  The models chosen are those discussed in Ref.\,\cite{Yan:2018bez}, which range from (\emph{a}) standard functions to (\emph{b}) parametrisations of experimental data and, finally, (\emph{c}) a theoretical calculation.
\begin{figure*}[!t]
	\renewcommand{\thefigure}{S1}
	\includegraphics[scale=0.54]{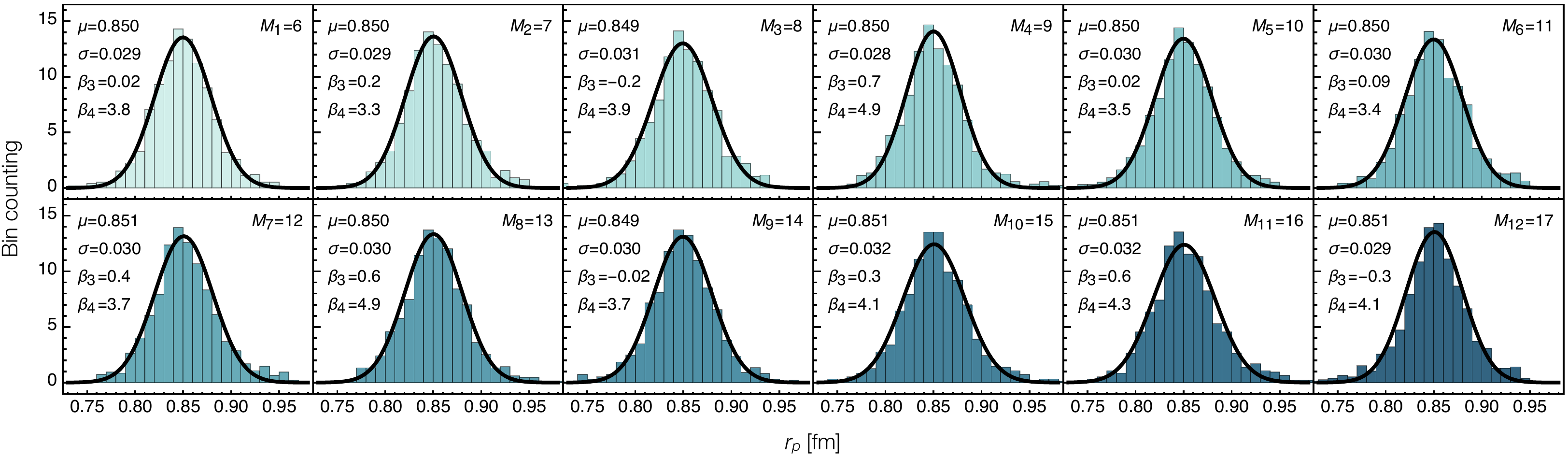}
	\caption{\label{fig:M-independence}
Probability distribution function associated with the SPM extraction of $r_p$ from $1\,000$ replicas generated using a dipole model, Eq.\,\eqref{dipole}, with $r_p=0.85\,$fm, from PRad data at $1.1\,$GeV beam energy kinematics.  Plainly, the reconstructed probability distribution is a Gaussian whose characteristics are practically insensitive to the number of input points $M_j$.}
\end{figure*}

\smallskip

\noindent\emph{a}. \underline{Standard functions}.
In this category, we have considered \cite{Borkowski:1975ume}: a monopole form
\begin{align}
	G^p_E(Q^2)&=\frac1{1+Q^2/p_1};&
	p_1&=\frac{0.467255}{2r_p^2}\ {\rm GeV}^2,\tag{a.1} \label{mod1}
\end{align}
corresponding to a Yukawa charge distribution of the proton; a dipole form
\begin{align}
	G^p_E(Q^2)&=\frac1{(1+Q^2/p_1)^2};&
	p_1&=\frac{0.467255}{r_p^2}\ {\rm GeV}^2,\tag{a.2}
	\label{dipole}
\end{align}
corresponding to an exponential charge distribution; and, finally, a Gaussian form
\begin{align}
	G^p_E(Q^2)&=\exp\left(-\frac{Q^2}{p_1}\right);&
	p_1&=\frac{0.467255}{r_p^2}\ {\rm GeV}^2.\tag{a.3} \label{mod3}
\end{align}
For all these models, $r_p=0.85\,$fm.

\smallskip

\noindent{\emph{b}}. \underline{Parametrisations of experimental data}.
The first parametrisation considered is that in Ref.\,\cite{Kelly:2004hm}:
\begin{align}
	G^p_E(Q^2)&=\frac{1+a_1\tau}{1+b_1\tau+b_2\tau^2+b3\tau^3};&
	\tau&=\frac{Q^2}{4m_p^2},\tag{b.1} \label{EqKelly}
\end{align}
with $m_p=0.938272\,$GeV being the proton mass. The coefficients $a_i$, $b_i$, and the expected value of $r_p$ are:
\begin{align}
	&\begin{tabular}{c|c|c|c}
		\hline
		$a_1$ & $b_1$ & $b_2$ & $b_3$ \\
		\hline
		-0.24 & 10.98 & 12.82 & 21.97
	\end{tabular}\,;&
	r_p&=0.8277\ {\rm fm}.
	\nonumber
\end{align}

Two different fits are employed in Refs.\,\cite{Arrington:2003qk, Arrington:2006hm}.  In the first \cite{Arrington:2003qk}, the proton's form factor is expressed via the following series:
\begin{align}
	G^p_E(Q^2)&=\left(1+\sum_{i=1}^6p_{2i}Q^{2i}\right)^{-1},\tag{b.2} \label{EqArrington1}
\end{align}
with $r_p=0.8681\,$fm and (in GeV$^{-2i}$)
\begin{align}
	\begin{tabular}{c|c|c|c|c|c}
	\hline
	$p_2$ & $p_4$ & $p_6$ & $p_8$ & $p_{10}$ & $p_{12}$ \\
	\hline
	3.226 & 1.508 & -0.3773 & 0.611 & -0.1853 & 1.596$\times 10^{-2}$
	\end{tabular}\, .
	\tag{b.3}
\end{align}

The second \cite{Arrington:2006hm} implements $r_p=0.8965\,$fm and is characterised by a fifth-order continuous fraction:
\begin{equation}
%
G^p_E(Q^2) = \frac{1}
{1+
\frac{p_1 Q^2}{1+\frac{p_2 Q^2}{1+ \frac{p_3 Q^2}{1+\frac{p_4 Q^2}{1+ p_{5} Q^2}}}}}\,,
\tag{b.4}
\label{EqArrington2}
\end{equation}
with coefficients (in GeV$^{-2}$)
\begin{align}
	\begin{tabular}{c|c|c|c|c}
	\hline
		$p_1$ & $p_2$ & $p_3$ & $p_4$ & $p_5$ \\
		\hline
		3.440 & -0.178 & -1.212 & 1.176 & -0.284
	\end{tabular}\, . \tag{b.5}
\end{align}

Ref.\,\cite{Bernauer:2013tpr} employed a tenth-order polynomial expansion of $G_E^p$:
\begin{align}
	G^p_E(Q^2)&=1+\sum_{i=1}^{10}p_iQ^{2i},\tag{b.6}
	\label{Bernauer2014}
\end{align}
with proton radius $r_p=0.8871\,$fm and the following coefficients (in GeV$^{-i}$)
\begin{equation}
	\begin{tabular}{c|c|c|c|c} \hline
		$p_1$ & $p_2$ & $p_3$ & $p_4$ & $p_5$ \\\hline
		-3.3686&  14.5606&  -88.1912&  453.6244&  -1638.7911 \\ \hline
    $p_6$ & $p_7$ & $p_8$ & $p_9$ & $p_{10}$\\\hline
    3980.7174& -6312.6333&  6222.3646&  -3443.2251&  814.4112
	\end{tabular}\,. \tag{b.7}
\end{equation}

The final parametrisation is that in Ref.\,\cite{Ye:2017gyb}, which is a fit to the world's data, yielding $r_p= 0.879\,$fm, expressing the proton's electric form factor as
\begin{equation}
	G^p_E(Q^2) =1+\sum_{i=1}^{13}a_iz^i(Q^2),\tag{b.8}
	\label{Ye2018}
\end{equation}
where
\begin{equation} z(Q^2) =\frac{\sqrt{t_\mathrm{cut}+Q^2}-\sqrt{t_\mathrm{cut}-t_0}} {\sqrt{t_\mathrm{cut}+Q^2}+\sqrt{t_\mathrm{cut}-t_0}},\tag{b.9}
\end{equation}
with $t_\mathrm{cut}=4m^2_\pi$, the two-pion particle production threshold ($m_\pi=0.13957\,$GeV is the pion mass), and $t_0=0.7\,$GeV.  The coefficients are:
\begin{equation}
\begin{array}{ll}
a_1 =\phantom{-}0.239163298067\,, & a_2 = -1.10985857441\,,\\
a_3 = \phantom{-}1.44438081306\,, & a_4 = \phantom{-}0.479569465603\,,\\
a_5 = -2.28689474187\,, & a_6 = \phantom{-}1.12663298498\,, \\
a_7 = \phantom{-}1.25061984354 \,, & a_8 = -3.63102047159\,,\\
a_9 = \phantom{-}4.08221702379 \,, & a_{10} = \phantom{-}0.504097346499\,, \\
a_{11} =  -5.08512046051 \,, & a_{12} = \phantom{-}3.96774254395\,, \\
a_{13} = -0.981529071103\,. &
\end{array}\tag{b.10}
\end{equation}

\smallskip

\noindent{\emph{c}}. \underline{Theoretical calculation}.
Ref.\,\cite{Alarcon:2018zbz} exploited a form factor analysis method that uses a combination of effective field theory and dispersion relations.  A parametrisation of the results, yielding the CODATA value of the proton radius \cite{Mohr:2015ccw}: $0.8765\,$fm, is given by
%
\begin{equation}
G^p_E(Q^2)=\frac{1+a_2Q^2+a_4Q^4+a_6Q^6}{1+b_2Q^2+b_4Q^4+b_6Q^6+b_8Q^8}
\tag{c.1}, \label{EqWeiss}
\end{equation}
with the coefficients (in GeV$^{-2i}$)
\begin{align}
	& \begin{tabular}{c|c|c}	\hline
		$a_2$ & $a_4$ & $a_6$ \\\hline
		$6.57333$ & $-6.63059$ & $3.97691$
	\end{tabular}\, , & \nonumber
\\[-2ex]
& \tag{c.2} & \\
	& \begin{tabular}{c|c|c|c}		\hline
		$b_2$ & $b_4$ & $b_6$ & $b_8$ \\
		\hline
		9.86132 & 16.8718 & 1.47422 & 0.106822
	\end{tabular}\, .&  \nonumber
\end{align}

\smallskip

\noindent{\it SM\,II. Validation results.} ---
For each of the models described and experimental data sets used:
(\emph{i}) we generate $10^3$ replicas for a fixed radius value, $r^*_p$;
(\emph{ii}) smooth each replica with the associated optimal parameter;
(\emph{iii}) use the SPM to obtain $r^{M_j}_p$ and $\sigma^{M_j}_p$, varying the number of input points $\{M_j=5+j\,\vert\ j = 1,\dots,n_{\s{M}};\,n_{\s{M}}= 12\}$;
and
(\emph{iv}) calculate the final SPM result using Eq.\,(4)\,--\,main text and compare it with the input value.

\begin{figure}[!t]
	\renewcommand{\thefigure}{S2}
	\includegraphics[scale=.6]{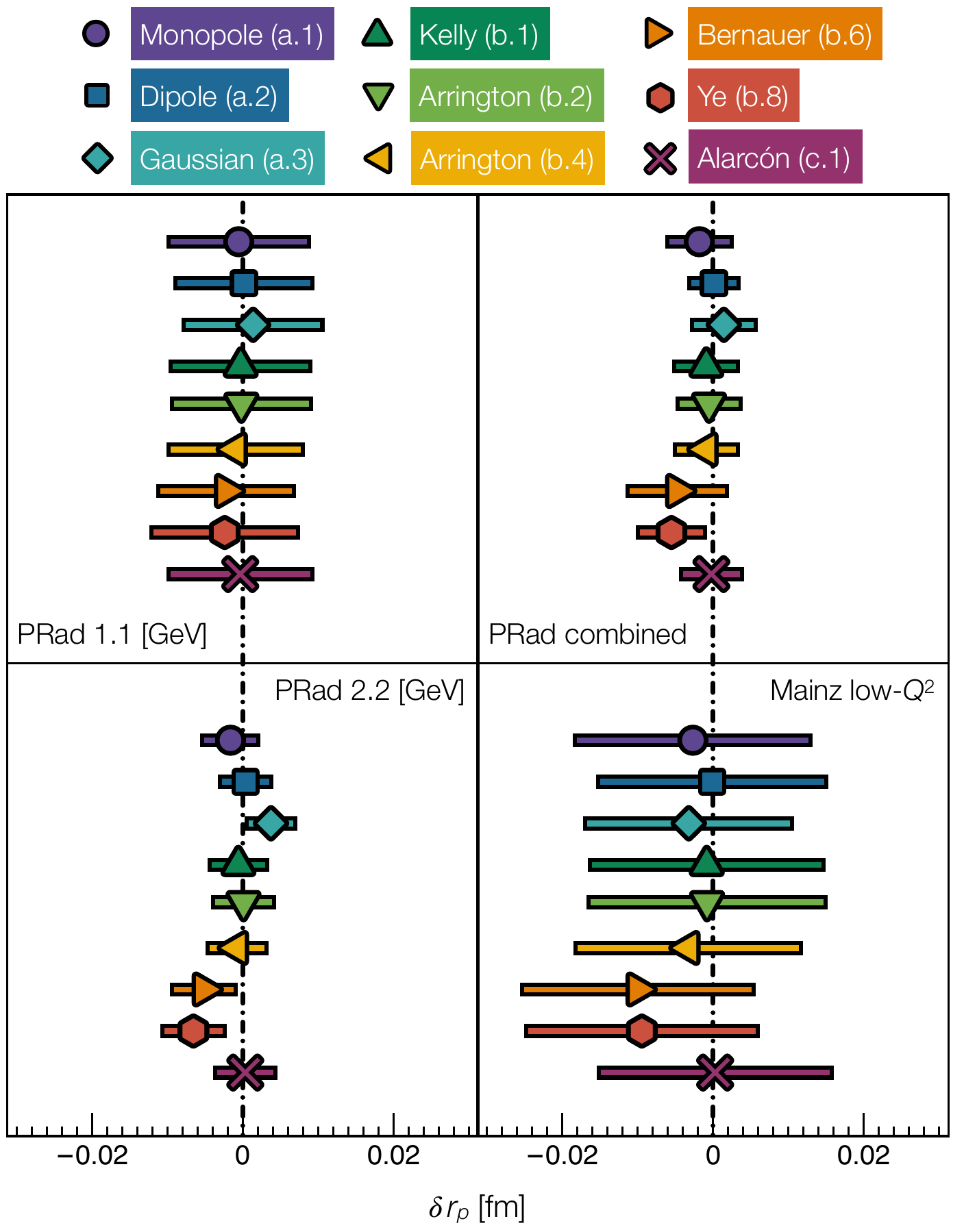}
	\caption{\label{fig:bias-vs-model} 
    Bias, $\delta r_p$, and associated standard error, $\sigma_r$, for the SPM extrapolations of the proton radius from $10^3$ replicas generated using the models described in Sec.\,SM\,I -- Eqs.\,\eqref{mod1}\,-\,\eqref{mod3};
    Eqs.\,\eqref{EqKelly}, \eqref{EqArrington1}, \eqref{EqArrington2}, \eqref{Bernauer2014}, \eqref{Ye2018};
    and Eq.\,\eqref{EqWeiss} --
    for the different $Q^2$ ranges corresponding to: PRad at $1.1\,$GeV beam energy (upper-left), $2.2\,$GeV (lower-left), combined data sets (upper-right) \cite{Xiong:2019umf}; and the low-$Q^2$ Mainz data set (lower-right) \cite{Bernauer:2010wm}.  In almost all cases, the SPM extrapolations are robust ($|\delta r_p|<\sigma_r$).  For the PRad data at $2.2\,$GeV and combined kinematics, the SPM results for the Eq.\,\eqref{Bernauer2014} and \eqref{Ye2018} generators are marginally robust ($|\delta r_p|\sim\sigma_r$).}
\end{figure}

\textbf{Observation~1}.
For a given $M_j$, the distribution of SPM extracted radii is a Gaussian whose characteristics are practically independent of $M_j$.  This is illustrated in Fig.\,\ref{fig:M-independence} for the case of replicas generated using the dipole functional form, Eq.\,\eqref{dipole}, with $r^\ast_p=0.85\,$fm and the PRad kinematics at $1.1\,$GeV beam energy.
Qualitatively identical behaviour is found in all other cases considered.
For this generator and data set, the SPM yields $r_p=0.8501\,$fm and
\begin{equation}
	\Bigg[\frac1{n^2_{\s M}}
    \sum_{j=1}^{n_M}(\sigma^{\s{M_j}}_r)^2\Bigg]^{\frac12}=0.0088\,{\rm fm}\,,\;
	 \sigma_{\s{\delta\!M}}=0.0004\,{\rm fm.} \tag{II.1}
\end{equation}
As anticipated, $\sigma_{\s{\delta\!M}}\ll \sigma^{\s{M_j}}_r$.

\textbf{Observation~2}.
Defining the bias as \mbox{$\delta r_p=r_p-r^*_p$}, then the SPM extraction of the proton radius is robust: in almost all cases, $|\delta r_p| < \sigma_r$, where $\sigma_r$ is the standard error in Eq.\,(4)\,--\,main text.
This is demonstrated by Fig.\,\ref{fig:bias-vs-model}, which displays $\delta r_p$ as obtained using the SPM to extract $r_p$ from all nine generators described in Sec.\,SM\,I over four different $Q^2$ ranges and statistical errors: PRad at $1.1\,$GeV beam energy,
$2.2\,$GeV beam, $1.1$, $2.2\,$GeV combined \cite{Xiong:2019umf}; and the low-$Q^2$ Mainz data \cite{Bernauer:2010wm}.
In all but 3 of the 36 cases, $|\delta r_p| < \sigma_r$.
The three exceptions are PRad $2.2\,$GeV kinematics with the generators in Eqs.\,\eqref{Bernauer2014}, \eqref{Ye2018}, and the PRad-combined with the Eq\,\eqref{Ye2018} generator.
This outcome is consistent with the analysis in Ref.\,\cite{Yan:2018bez}, which reported that the Eq.\,\eqref{Bernauer2014}, \eqref{Ye2018} generators consistently showed the highest bias of all fitters, independent of the $Q^2$-binning (see Fig.\,11 therein).
In this connection, it is also notable that, as already remarked, the PRad $2.2\,$GeV error is much smaller than that associated with the $1.1\,$GeV data, driving down the error in the combined set to roughly one-half of that found with the $1.1\,$GeV data alone.

\begin{figure*}[!t]
	\renewcommand{\thefigure}{S3}
	\includegraphics[scale=0.61]{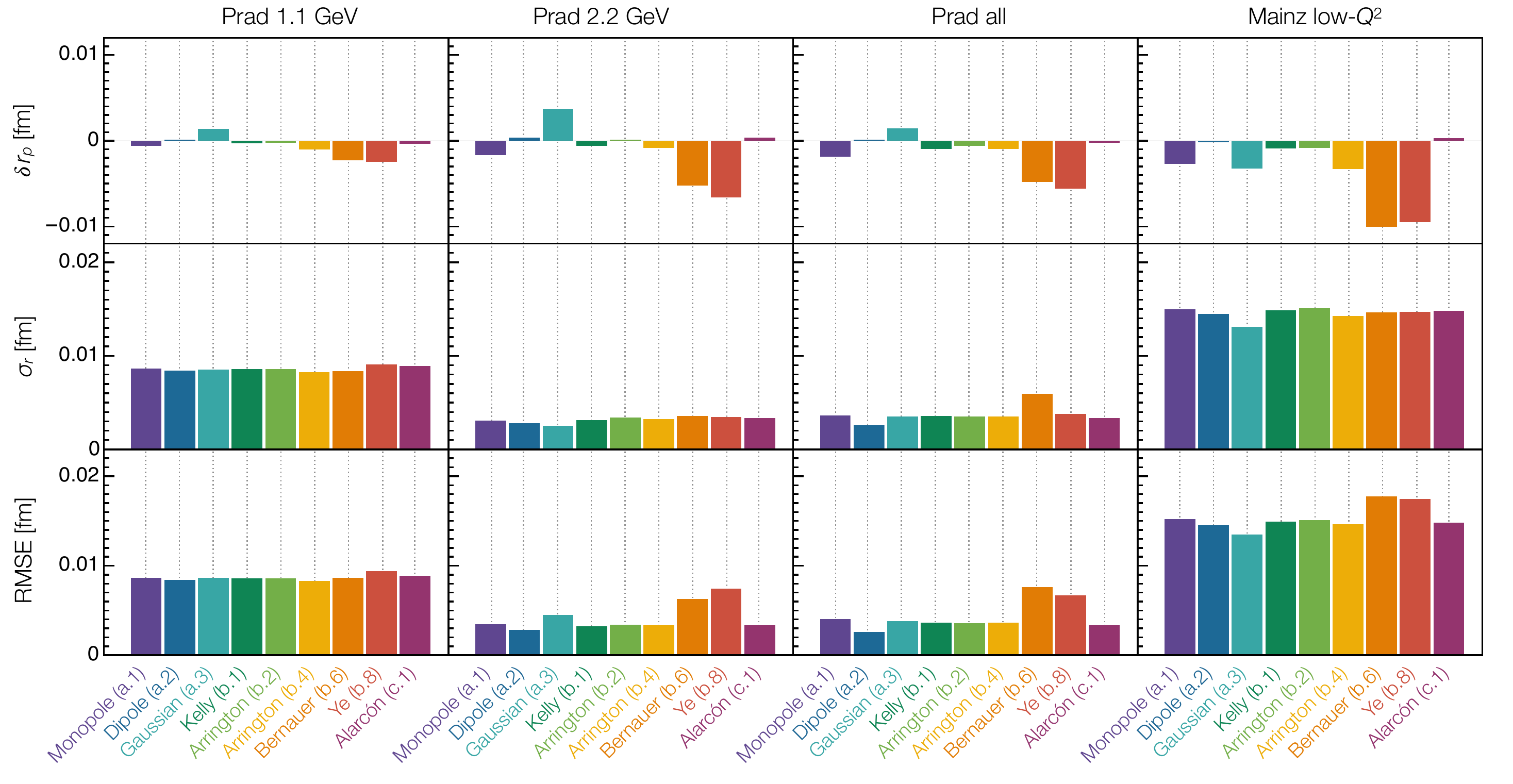}
	\caption{\label{fig:Delta-r_sigma_RMSE} 
Bias, $\delta r_p$, standard error, $\sigma_r$, and RMSE, Eq.\,\eqref{EqRMSE}, for SPM extrapolations of the proton radius from $10^3$ replicas generated using the nine models described in Sec.\,SM\,I.  Notably, within a given experimental data set, the RMSE is approximately independent of the generator used.}
\end{figure*}

\textbf{Observation~3}.
Defining the root mean square error (RMSE)
\begin{equation}
	\mathrm{RMSE}=\sqrt{(\delta r_p)^2+\sigma_r^2}\,, \tag{II.2}
\label{EqRMSE}
\end{equation}
then for any given experimental data set, as shown in Fig.\,\ref{fig:Delta-r_sigma_RMSE}, the SPM analysis produces RMSE values that are approximatively independent of the generator used to obtain the replicas.  This means that the SPM procedure satisfies a standard ``goodness of fit'' criterion \cite{Yan:2018bez}, in consequence of which our SPM extractions of the proton radius can objectively be judged as sound.



\begin{thebibliography}{10}

\bibitem{Marciano:1979wa}
W.~J. Marciano and H.~Pagels,
\newblock Nature {\bf 279}, 479 (1979).

\bibitem{Durr:2008zz}
S.~Durr {\em et~al.},
\newblock Science {\bf 322}, 1224 (2008).

\bibitem{Eichmann:2016yit}
G.~Eichmann, H.~Sanchis-Alepuz, R.~Williams, R.~Alkofer and C.~S. Fischer,
\newblock Prog. Part. Nucl. Phys. {\bf 91}, 1 (2016).

\bibitem{Qin:2019hgk}
S.-X. Qin, C.~D. Roberts and S.~M. Schmidt,
\newblock Few Body Syst. {\bf 60}, 26 (2019).

\bibitem{millennium:2006}
{\emph{The Millenium Prize Problems}, eds. J. Carlson, A. Jaffe, and A. Wiles.
  (American Mathematical Society, Providence, 2006)}.

\bibitem{Pohl:2010zza}
R.~Pohl {\em et~al.},
\newblock Nature {\bf 466}, 213 (2010).

\bibitem{Pohl:2013yb}
R.~Pohl, R.~Gilman, G.~A. Miller and K.~Pachucki,
\newblock Ann. Rev. Nucl. Part. Sci. {\bf 63}, 175 (2013).

\bibitem{Carlson:2015jba}
C.~E. Carlson,
\newblock Prog. Part. Nucl. Phys. {\bf 82}, 59 (2015).

\bibitem{Mohr:2015ccw}
P.~J. Mohr, D.~B. Newell and B.~N. Taylor,
\newblock Rev. Mod. Phys. {\bf 88}, 035009 (2016).

\bibitem{Antognini:1900ns}
A.~Antognini {\em et~al.},
\newblock Science {\bf 339}, 417 (2013).

\bibitem{Beyer:2017gug}
A.~Beyer {\em et~al.},
\newblock Science {\bf 358}, 79 (2017).

\bibitem{Fleurbaey:2018fih}
H.~Fleurbaey {\em et~al.},
\newblock Phys. Rev. Lett. {\bf 120}, 183001 (2018).

\bibitem{Bernauer:2010wm}
J.~C. Bernauer {\em et~al.},
\newblock Phys. Rev. Lett. {\bf 105}, 242001 (2010).

\bibitem{Xiong:2019umf}
W.~Xiong {\em et~al.},
\newblock Nature {\bf 575}, 147 (2019).

\bibitem{Bezginov1007}
N.~Bezginov {\em et~al.},
\newblock Science {\bf 365}, 1007 (2019).

\bibitem{Grinin1061}
A.~Grinin {\em et~al.},
\newblock Science {\bf 370}, 1061 (2020).

\bibitem{Pohl1:2016xoo}
R.~Pohl {\em et~al.},
\newblock Science {\bf 353}, 669 (2016).

\bibitem{PhysRev.167.1411}
L.~Schlessinger,
\newblock Phys. Rev. {\bf 167}, 1411 (1968).

\bibitem{Schlessinger:1966zz}
L.~Schlessinger and C.~Schwartz,
\newblock Phys. Rev. Lett. {\bf 16}, 1173 (1966).

\bibitem{Tripolt:2016cya}
R.~A. Tripolt, I.~Haritan, J.~Wambach and N.~Moiseyev,
\newblock Phys. Lett. B {\bf 774}, 411 (2017).

\bibitem{Gilman:2013vma}
R.~Gilman,
\newblock AIP Conf. Proc. {\bf 1563}, 167 (2013).

\bibitem{Gasparian:2014rna}
A.~Gasparian,
\newblock EPJ Web Conf. {\bf 73}, 07006 (2014).

\bibitem{Kraus:2014qua}
E.~Kraus, K.~E. Mesick, A.~White, R.~Gilman and S.~Strauch,
\newblock Phys. Rev. C {\bf 90}, 045206 (2014).

\bibitem{Lorenz:2014vha}
I.~T. Lorenz and U.-G. Mei{\ss}ner,
\newblock Phys. Lett. B {\bf 737}, 57 (2014).

\bibitem{Griffioen:2015hta}
K.~Griffioen, C.~Carlson and S.~Maddox,
\newblock Phys. Rev. C {\bf 93}, 065207 (2016).

\bibitem{Higinbotham:2015rja}
D.~W. Higinbotham {\em et~al.},
\newblock Phys. Rev. C {\bf 93}, 055207 (2016).

\bibitem{Hayward:2018qij}
T.~B. Hayward and K.~A. Griffioen,
\newblock Nucl. Phys. A {\bf 999}, 121767 (2020).

\bibitem{Zhou:2018bon}
S.~Zhou, P.~Giulani, J.~Piekarewicz, A.~Bhattacharya and D.~Pati,
\newblock Phys. Rev. C {\bf 99}, 055202 (2019).

\bibitem{Alarcon:2018zbz}
J.~M. Alarc{\'o}n, D.~W. Higinbotham, C.~Weiss and Z.~Ye,
\newblock Phys. Rev. C {\bf 99}, 044303 (2019).

\bibitem{Higinbotham:2019jzd}
S.~K. Barcus, D.~W. Higinbotham and R.~E. McClellan,
\newblock Phys. Rev. C {\bf 102}, 015205 (2020).

\bibitem{Hammer:2019uab}
H.-W. Hammer and U.-G. Mei{\ss}ner,
\newblock Sci. Bull. {\bf 65}, 257 (2020).

\bibitem{Yan:2018bez}
X.~Yan {\em et~al.},
\newblock Phys. Rev. C {\bf 98}, 025204 (2018).

\bibitem{Chen:2018nsg}
C.~Chen {\em et~al.},
\newblock Phys. Rev. D {\bf 99}, 034013 (2019).

\bibitem{Binosi:2018rht}
D.~Binosi {\em et~al.},
\newblock Phys. Lett. B {\bf 790}, 257 (2019).

\bibitem{Binosi:2019ecz}
D.~Binosi and R.-A. Tripolt,
\newblock Phys. Lett. B {\bf 801}, 135171 (2020).

\bibitem{Eichmann:2019dts}
G.~Eichmann, P.~Duarte, M.~Pe{\~n}a and A.~Stadler,
\newblock Phys. Rev. D {\bf 100}, 094001 (2019).

\bibitem{Yao:2020vef}
Z.-Q. Yao {\em et~al.},
\newblock Phys. Rev. D {\bf 102}, 014007 (2020).

\bibitem{10.5555/1403886}
W.~H. Press, S.~A. Teukolsky, W.~T. Vetterling and B.~P. Flannery,
\newblock {\em Numerical Recipes 3rd Edition: The Art of Scientific Computing},
  3 ed. (Cambridge University Press, USA, 2007).

\bibitem{Reinsch:1967aa}
C.~H. Reinsch,
\newblock Numer. Math. {\bf 10}, 177 (1967).

\bibitem{Craven:1978aa}
P.~Craven and G.~Wahba,
\newblock Numer. Math. {\bf 31}, 377 (1978).

\bibitem{Borkowski:1975ume}
F.~Borkowski, G.~G. Simon, V.~H. Walther and R.~D. Wendling,
\newblock Z. Phys. A {\bf 275}, 29 (1975).

\bibitem{Kelly:2004hm}
J.~J. Kelly,
\newblock Phys. Rev. C {\bf 70}, 068202 (2004).

\bibitem{Arrington:2003qk}
J.~Arrington,
\newblock Phys. Rev. C {\bf 69}, 022201 (2004).

\bibitem{Arrington:2006hm}
J.~Arrington and I.~Sick,
\newblock Phys. Rev. C {\bf 76}, 035201 (2007).

\bibitem{Bernauer:2013tpr}
J.~C. Bernauer {\em et~al.},
\newblock Phys. Rev. C {\bf 90}, 015206 (2014).

\bibitem{Ye:2017gyb}
Z.~Ye, J.~Arrington, R.~J. Hill and G.~Lee,
\newblock Phys. Lett. B {\bf 777}, 8 (2018).

\bibitem{Alarcon:2017ivh}
J.~M. Alarc\'on and C.~Weiss,
\newblock Phys. Rev. C {\bf 96}, 055206 (2017).

\end{thebibliography}

\end{document}